# Impact of Si nanocrystals in a-SiO$_x$<Er> in C-Band emission for applications in resonators structures

D.S.L Figueira[1,2], D. Mustafa[1], L.R. Tessler[1] and N.C. Frateschi[2,1]

[1]Instituto de Física "Gleb Wataghin", Universidade Estadual de Campinas, Unicamp, 13083-970, Campinas, São Paulo, Brasil. [2]Centro de Componentes Semicondutores, Universidade Estadual de Campinas, Unicamp, 13083-870, Campinas, São Paulo, Brasil

*Abstract* — Si nanocrystals (Si-NC) in a-SiO$_x$<Er> were created by high temperature annealing. Si-NC samples have large emission in a broadband region, 700nm to 1000nm. Annealing temperature, annealing time, substrate type, and erbium concentration is studied to allow emission at 1550 nm for samples with erbium. Emission in the "C-Band" region is largely reduced by the presence of Si-NC. This reduction may be due to less efficient energy transfer processes from the nanocrystals than from the amorphous matrix to the Er$^{3+}$ ions, perhaps due to the formation of more centro-symmetric Er$^{3+}$ sites at the nanocrystal surfaces or to very different optimal erbium concentrations between amorphous and crystallized samples.

*Index Terms* — Silicon photonics, nanocrystals, resonator.

## I. INTRODUCTION

The development of silicon based active optical devices has been of interest in recent years due the possibility of integration with CMOS technology. Silicon doping with rare-earth elements is an alternative for efficient light emission in the C-band.

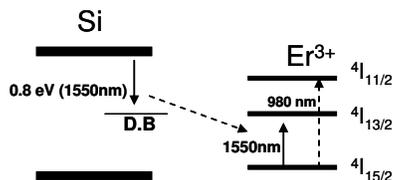

Fig. 1. Squematic energy diagram of a-SiO$_x$<Er> showed the energy separation between the Si conduction band and the dangling bond (DB), formed due insertion of Er$^{3+}$ in Si matrix, corresponds approximately to 0.8 eV is responsible for large emission these materials. The oxygen is responsible for detuning of DB separation for efficient emission in 0.8eV.

Rare-earth elements inserted in some semiconductors and insulators emit luminescence due to internal transitions of the incomplete *4f* levels in wavelengths that are almost independent of the matrix. The *f-f* transition is forbidden (because this, is very hard access this transition directly, a laser with bandwidth very narrow is required). Particularly, in amorphous silicon sub-oxide doped with erbium (a-Si<Er>), it is clear that transfer of energy from the matrix to Er$^{3+}$ ions occurs [1,2], allowing excitation from photons that effectively are absorbed by the a-Si:H. Some models attribute the good excitation efficiency of the Er in a-Si:H to a defect related Auger excitation, based on the fact that the Er$^{3+}$ transition energy corresponds approximately to the energy separation between the conduction band and the dangling bond energy [3], shown in figure 1.

Si-NC in SiO$_x$ matrices presents light emission in NIR and is interesting to access the $^4I_{13/2}$ to $^4I_{11/2}$ transition of erbium and improvement the efficient this materials for photonics applications [4].

In this work we present a method for fabrication Si-NC in a-SiO$_x$<Er> using RF co-sputtering and high temperature annealing. A direct application this material is fabrication of microdisk resonators with whispering gallery modes (WGM) [5], offering great advantage for stimulated light emission generation in small volumes where long photon lifetime is achieved with simple processing steps [5,6]. Also, light emission occurs mainly along the substrate plane being potentially suitable for photonic integration. Moreover, with the use of non-conventional structures such as stadium, one may achieve higher emission spectral and directional control, expanding the integration possibilities [7,8].

## II. EXPERIMENTAL PROCEDURES

Si type "n" <100> with 260 Ω.cm, Si type "n" <111> with 2000 Ω.cm, and SiO$_2$ (1.25μm) were used to grow a film of a-SiO$_x$<Er>. The SiO$_2$ substrate was prepared by wet oxidation in Si ("n" <100>) at 1100°C for 270 minutes in a flux of 1.0 l/min of O$_2$ and water vapor. The a-SiO$_x$<Er> thin films (0.3 μm) were prepared by RF sputtering co-deposition on Si/SiO$_2$ substrates. The base pressure of the vacuum chamber was 2x10$^6$ mbar and the sputtering was carried out in RF mode with the bias fixed at 1kV from circular sources in an atmosphere of 8x10$^{-3}$ mbar of argon. The oxygen content in the films corresponds to O$_2$ partial pressure during deposition in 5.5x10$^{-5}$ mbar. As a source, we used ultra pure crystalline Si wafer mixed with metallic erbium (0, 1, and 3 platelets of 4mm$^2$ each). The substrate temperature was maintained at 240°C and the deposition rate of roughly 0.5Å/s. Using this technique becomes possible to obtain a-Si:H film doped with Er$^{3+}$ without complicated steps.

The thermal treatment consisted of two cumulative annealing procedures. The first annealing was at 400°C for one hour to effuse Ar remaining in the film [9]. The second annealing was varying at 1100ºC to 1200ºC for 0.5 hour to 1.5 hour to study the mechanism of Si-NC formation [4].

Photoluminescence (PL) was measured in a standard setup using 514nm line of an $Ar^+$ laser (100mW at the sample) as excitation. Detection used either a 2.0 m spectrograph with $LN_2$ cooled Ge detector for C-Band region, and a Si detector for NIR region (700nm to1000nm). The measurements were realized at room temperature.

### III. RESULTS

The evidence of Si-NC formation is emission in NIR. Optimization of NIR emission was performed changing the annealing temperature, annealing time, $Er^{3+}$ concentration, and substrate. Best emission condition was observed for a 1 $Er^{3+}$ platelet at 1150ºC in 1h of annealing, as shown in fig.2.

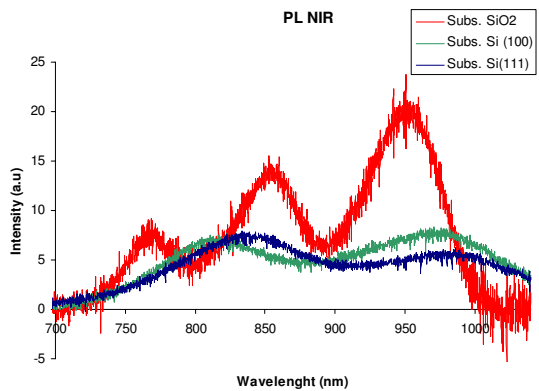

Fig. 2. Room temperature NIR photoluminescence with different substrates. Annealing temperature 1150ºC, time annealing 1 hour. $Er^{3+}$ concentration 1platelet of $4mm^2$ .

Figure 2 shows the typical Si-NC emission with formation of two groups of NC island. Fabry-perot modes (due 1550nm emission) were observed in $SiO_2$ substrate, with high intensity emission in 960nm. As shown in fig. 2 there is a strong influence of substrate in Si-NC formation. In Si type "n" <100> (260 Ω.cm) substrate occurs a blue shift emission because the diffusion of dopant of substrate.

For identical conditions of annealing temperature and annealing time, with improvement in $Er^{3+}$ concentration for 3 platelets ($4mm^2$ each), the PL spectra changes, as shown in figure 3.

The PL intensity showed in fig. 3 decrease 3x compared to the conditions used to generate the graphic showed in figure 2. The PL emission of Si <111> substrate is dramatically reduced, and the PL emission of Si<100> substrate presents coalescence of Si-NC as showed in fig. 2.

For $Er^{3+}$ concentrations over 3 platelets, the PL emission in NIR is inexistent. For samples without $Er^{3+}$ the PL emission is much reduced when compared to the PL spectra showed in fig. 2.

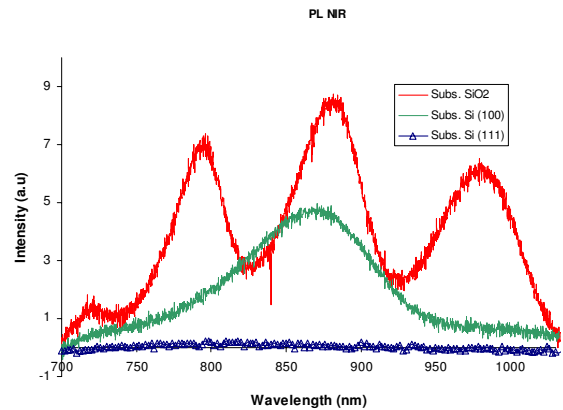

Fig. 3. Room temperature NIR photoluminescence with different substrates. Annealing temperature 1150ºC, time annealing 1 hour. $Er^{3+}$ concentration 3 platelets of $4mm^2$(each).

The PL emission in C band region is showed in figure 4.a for the same sample used in the best NIR emission. Without annealing, the sample has 7x more PL intensity, as showed in figure 4.b

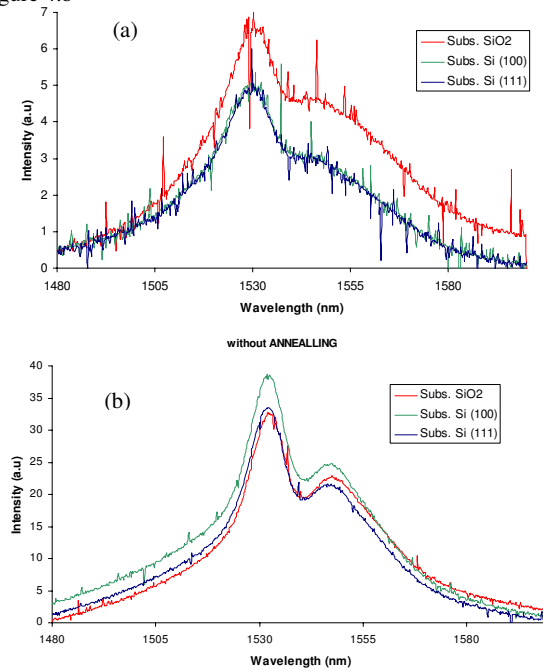

Fig.4. a)Room temperature C Band photoluminescence with different substrates. Annealing temperature 1150ºC, time annealing 1 hour. $Er^{3+}$ concentration 1platelet of 4mm$^2$ b) Room temperature C Band PL without anneling.

IV. CONCLUSION

This results show the impact in C band for Si-NC in SiO$_x$<Er>. A reduction of PL emission in 1550nm was measurement. However an improvement of this emission occurs in SiO$_2$ substrate, as showed in fig.4a.

As showed there are a strong dependence of Erbium and the formation of Si-NC in a-SiO$_x$<Er>.

Co-sputtering is a simple alternative process to the formation of Si-NC in SiO$_x$<Er>. However optical pumping process is not efficient method. Further studies will explore the preparation parameter for spectral emission engineering (centered in 980nm), and will include electroluminescence, HRTEM to access the nanocrystal formation and EXAFS to determine the evolution of the Er environment.

Microdisk and microstadium resonators will be fabricated to evaluate emission properties. Also, preliminary evaluation of optical injection on samples with Si-NC and $Er^{3+}$ will be presented.